\documentclass[aps,prb,superscriptaddress,twocolumn,showpacs]{revtex4}
\usepackage{amsmath,bm}
\usepackage{graphicx,epsfig,braket,amssymb,amsfonts,color}

\newcommand{\ba}{\begin{eqnarray}}
\newcommand{\ea}{\end{eqnarray}}
\newcommand{\be}{\begin{equation}}
\newcommand{\ee}{\end{equation}}
\newcommand{\bd}{\begin{displaymath}}
\newcommand{\ed}{\end{displaymath}}
\renewcommand{\v}[1]{{\bf #1}}
\newcommand{\bpm}{\begin{pmatrix}}
\newcommand{\epm}{\end{pmatrix}}
\newcommand{\nn}{\nonumber \\}

\begin{document}

\preprint{APS/123-QED}

\title{Interfacial Rashba magnetoresistance of two-dimensional electron gas at LaAlO$_3$/SrTiO$_3$ interface}

\author{Kulothungasagaran Narayanapillai}
\thanks{These two authors contributed equally to this work.}
\affiliation{Department of Electrical and Computer Engineering, NUSNNI, National University of Singapore, 117576, Singapore}

\author{Gyungchoon Go}
\thanks{These two authors contributed equally to this work.}
\affiliation{Department of Materials Science and Engineering, Korea University, Seoul 02841, Korea}

\author{Rajagopalan Ramaswamy}
\affiliation{Department of Electrical and Computer Engineering, NUSNNI, National University of Singapore, 117576, Singapore}

\author{Kalon Gopinadhan}
\affiliation{Department of Electrical and Computer Engineering, NUSNNI, National University of Singapore, 117576, Singapore}

\author{Dongwook Go}
\affiliation{PCTP and Department of Physics, Pohang University of Science and Technology, Pohang 37673, Korea}

\author{Hyun-Woo Lee}
\affiliation{PCTP and Department of Physics, Pohang University of Science and Technology, Pohang 37673, Korea}

\author{Thirumalai Venkatesan}
\affiliation{Department of Electrical and Computer Engineering, NUSNNI, National University of Singapore, 117576, Singapore}
\affiliation{Department of Physics, National University of Singapore, Singapore, 117542, Singapore}

\author{Kyung-Jin Lee}
\email{kj_lee@korea.ac.kr}
\affiliation{Department of Materials Science and Engineering, Korea University, Seoul 02841, Korea}
\affiliation{KU-KIST Graduate School of Converging Science and Technology, Korea University, Seoul 02841, Korea}

\author{Hyunsoo Yang}
\email{eleyang@nus.edu.sg}
\affiliation{Department of Electrical and Computer Engineering, NUSNNI, National University of Singapore, 117576, Singapore}

\date{\today}

\begin{abstract}
We report the angular dependence of magnetoresistance in two-dimensional electron gas at LaAlO$_3$/SrTiO$_3$ interface. We find that this interfacial magnetoresistance exhibits a similar angular dependence to the spin Hall magnetoresistance observed in ferromagnet/heavy metal bilayers, which has been so far discussed in the framework of bulk spin Hall effect of heavy metal layer. The observed magnetoresistance is in qualitative agreement with theoretical model calculation including both Rashba spin-orbit coupling and exchange interaction.
Our result suggests that magnetic interfaces subject to spin-orbit coupling can generate a nonnegligible contribution to the spin Hall magnetoresistance and the interfacial spin-orbit coupling effect is therefore key to the understanding of various spin-orbit-coupling-related phenomena in magnetic/non-magnetic bilayers.
\end{abstract}

\pacs{85.75.-d; 73.20.-r; 75.47.-m; 75.70.Tj}

\maketitle

\section{Introduction}

Recent advances in deposition techniques enable film growth control at the molecular level with atomic precision. These advances allow us to study exotic oxide materials with interesting properties. Since the discovery of two-dimensional electron gas (2DEG) formation at the interface of two insulating materials, LaAlO$_3$ (LAO) and SrTiO$_3$ (STO)~\cite{Ohtomo}, the LAO/STO system has emerged as one of the central material systems in the oxide community as it exhibits intriguing properties including superconductivity~\cite{Reyren,Caviglia2008} and ferromagnetism~\cite{Brinkman,Ariando,Kalisky,Lee}. The ferromagnetism, which occurs on Ti site at the interface, has been evidenced by various measurement techniques such as scanning superconducting quantum interference device (SQUID)~\cite{Kalisky,Bert}, X-ray magnetic circular dichroism~\cite{Lee}, torque magnetometry~\cite{Li}, and magneto-transport~\cite{Joshua,Annadi,Narayanapillai,Fete, Diez, Caviglia2010, Hurand} driven by the Rashba spin-orbit coupling (SOC). Furthermore, the broken inversion symmetry at the interface results in the Rashba SOC which can be tuned by a gate voltage~\cite{Caviglia2010,Hurand}. Therefore, the LAO/STO interface is subject to both interfacial Rashba SOC and exchange coupling.

The magnetoresistance (MR) is a fundamental means to investigate charge and spin transport in condensed matter systems. In magnetic systems  in the presence of SOC, the longitudinal resistance depends on the magnetization direction, e.g. anisotropic MR~\cite{Thomson}. Recently, another type of angle-dependent MR, called spin Hall MR, was observed in ferromagnet (FM)/heavy metal (HM) bilayers~\cite{Nakayama2013,Cho,Kim2016}. Including both anisotropic MR and spin Hall MR, the longitudinal resistivity is given as
\begin{equation}\label{eq1}
\rho = \rho_0 + \Delta \rho_{1} m_x^2 - \Delta \rho_{2} m_y^2,
\end{equation}
where $\rho_0$ is the magnetization-direction-independent resistivity, $\Delta \rho_{1}$ ($\Delta \rho_{2}$) is the longitudinal resistivity change due to the anisotropic MR (spin Hall MR), and $m_x$ ($m_y$) is the normalized magnetization component longitudinal (transverse) to the current-flow direction in the film plane. The original theory~\cite{Chen} described the spin Hall MR as a consequence of bulk spin Hall effect (SHE) in HM by assuming no interfacial SOC effect. Recent theories however predicted an important role of the interfacial SOC at the FM/HM interface in the spin Hall MR~\cite{Grigoryan,Zhang}.

Furthermore, several experiments on FM/HM bilayers suggested that the role of the interfacial SOC should be carefully examined. For example, it was reported for spin pumping and inverse SHE experiments on Co/Pt bilayers~\cite{Rojas2} that the total dissipated transverse spin current from Co layer (measured through the effective damping) is substantially different from the spin current absorbed in the bulk part of Pt layer (measured through inverse SHE). This difference was ascribed to the spin memory loss~\cite{Kurt} describing the spin-flipping due to SOC at the Co/Pt interface~\cite{yen}. There were also spin-orbit torque experiments that cannot be explained by the bulk SHE mechanism alone but require an essential role of the interfacial SOC effect~\cite{Fan,Kurebayashi,Qiu,Oh}. A recent experiment~\cite{Berger} also found a close correlation, predicted for the interfacial SOC~\cite{KLLS}, between the Dzyaloshinskii-Moriya interaction and fieldlike spin-orbit torque~\cite{KLLS}.

Recent theories also suggested that the interfacial SOC effect is important for various SOC-related phenomena in FM/HM bilayers. First-principles approach~\cite{Freimuth,Wang} showed that both SHE and inverse SHE are largely enhanced at the FM/HM interface, i.e., a manifestation of the interfacial SOC effect. Boltzmann transport calculations also suggested the importance of the interfacial SOC effect for various spin transport phenomena~\cite{Haney,Amin1,Amin2}.

\begin{figure}[ttbp]
\includegraphics[width=80mm]{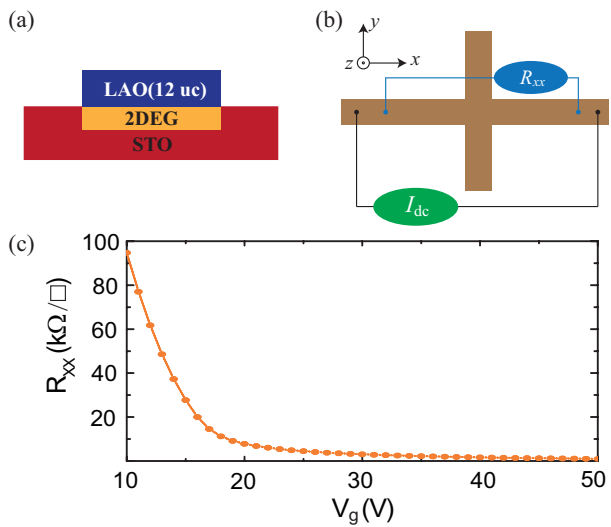}
\caption{(color online) (a)  Schematics of LAO/STO stack layers, (b)  Hall bar device with a measurement schematic for $R_{xx}$, and (c) The sheet resistance, $R_{xx}$, as a function of gate voltage $V_g$.
At low gate voltages, the sample is insulating and the resistance monotonically decreases with increasing $V_g$.}\label{fig:1}
\end{figure}

Consequently, it is natural to raise a question about the role of the interfacial SOC in the spin Hall MR. Answering this question is of critical importance not only for understanding the underlying physics of SOC-related phenomena in FM/HM bilayers but also for enhancing the SOC effects for applications.
For this purpose, we investigate the angular dependence of MR in 2DEG formed at the LAO/STO interface which has interfacial Rashba SOC and ferromagnetism. Unlike FM/HM bilayer structures where both bulk and interfacial SOC contributions coexist and it is therefore hard to differentiate one from the other, the LAO/STO 2DEG is an ideal system to investigate a pure interface effect because it is a single interface just like the FM/HM interface but does not have the bulk SHE.

\section{Experiments}

The devices were prepared as follows. As shown in Fig. \ref{fig:1}(a), LAO (12 unit cells; 1 u.c.= 0.379 nm) was grown on a TiO$_2$ terminated atomically smooth STO (001) single crystalline substrate, which was pre-treated with buffered oxide etch and air-annealed at 950~$^\circ$C for 1.5 h, by pulsed laser deposition (PLD) at 750~$^\circ$C in an oxygen partial pressure of 1 mTorr. The growth was monitored by in-situ reflective high energy electron diffraction. The sample was post-annealed at 750~$^\circ$C in the presence of oxygen in order to remove oxygen vacancies. Electron beam lithography was utilized to define the Hall bar structure using a negative tone resist. A blanket AlN$_x$ was deposited by PLD at room temperature followed by a lift-off process. Thereby, a 2DEG LAO/STO interface was formed only at the defined device structure. The Hall bar channels were designed parallel to the sample edge to align the channels along the crystallographic axis (001).

A measurement scheme for longitudinal resistance ($R_{xx}$) is shown in Fig. \ref{fig:1}(b). The electrical transport characterizations were performed in a physical property measurement system with a rotator used for angle-dependent MR measurements. We performed MR measurements as a function of various parameters including magnetic field $B_{\rm ext}$, gate voltage $V_g$, and rotation plane of magnetic field. For the gating, a back gate voltage ($V_g$) was applied through STO as the dielectric and silver as the back gate contacts.

In Fig. \ref{fig:1}(c), we depict a typical $R_{xx}$ versus $V_g$ curve at the temperature $T$ of 2$\,$K. We observe a monotonic decrease in $R_{xx}$ with increasing $V_g$, showing that the device characteristic changes from insulating to conducting as $V_g$ increases. Figure 2 shows the perpendicular magnetic field ($B_{\rm ext}$) dependence of $R_{xx}$ at various gate voltages. At low gate voltages ($V_g$ $<$ 40 V), $R_{xx}$ decreases with increasing $|B_{\rm ext}|$, which results from the weak localization (WL). At high gate voltages ($V_g$ $>$ 45 V), on the other hand, $R_{xx}$ increases with increasing $|B_{\rm ext}|$,  which implies that the WL correction is sub-dominant in conducting regimes.


\begin{figure}[ttbp]
\includegraphics[width=85mm]{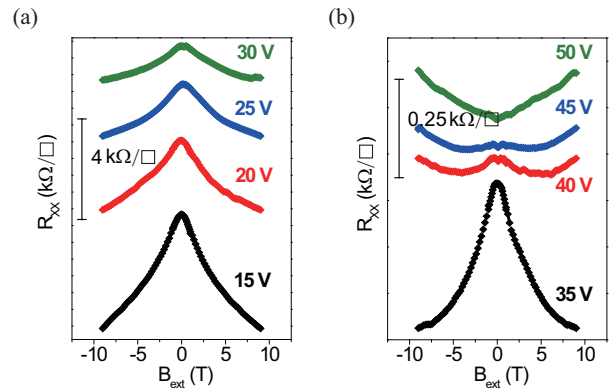}
\caption{(color online) Gate voltage dependence of $R_{xx}$.
Perpendicular magnetic field dependence of $R_{xx}$ for (a) $V_g$ = 15, 20, 25 and 30 V, and (b) $V_g$ = 35, 40, 45 and 50 V.}\label{fig:2}
\end{figure}

Figure 3 shows a representative result of the angle-dependent $R_{xx}$ for three rotation angles of $B_{\rm ext}$, i.e., $\alpha$-rotation in the $xy$-plane, $\beta$-rotation in the $yz$-plane, and $\gamma$-rotation in the $zx$-plane (see Fig. 3 for the definition of the angles). Here the current is always applied in the $x$-direciton. We observe that the normalized MR [$\equiv \tilde R_{xx}=(R_{xx}^{\rm max}-R_{xx}^{\rm min})/R_{xx}^{\rm min}$] is about $7\%$ in the $\alpha$- and $\beta$-rotations, while it is about $3\%$ in the $\gamma$-rotation~\cite{note1}.
An important observation is that the angle-dependent change in $\tilde R_{xx}$ is nonzero for the $\beta$-rotation.
As the LAO/STO interface has no contribution from the bulk SHE, this noticeable change in $\tilde R_{xx}$ for the $\beta$-rotation proves that $\Delta \rho_2$ term of Eq.~\eqref{eq1} is nonzero for the LAO/STO interface even without the bulk SHE. This angle-dependent MR at the LAO/STO interface can be named the interfacial Rashba MR as it originates from the Rashba SOC at the interface.
Futhermore, it is noteworthy that the interfacial Rashba MR is much larger than the reported spin Hall MR of FM/HM structures (0.01$-$1$\%$)~\cite{Nakayama2013,Cho,Kim2016}.
\begin{figure}[ttbp]
\includegraphics[width=85mm]{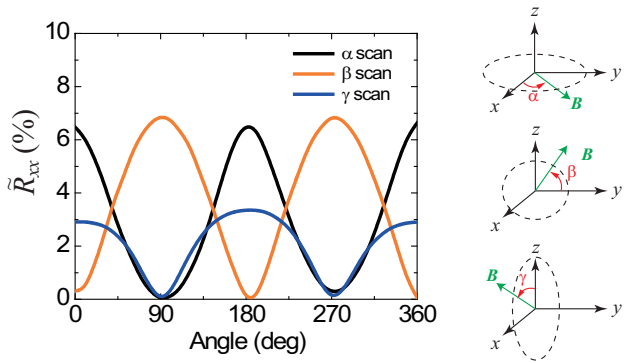}
\caption{(color online) Measured angular dependence of the normalized MR ($\tilde R_{xx}$) as a function of the rotating angle ($\alpha$, $\beta$, and $\gamma$) with applied field of $B_{\rm ext}$ = 9 T and gate voltage $V_g$ = 45 V. }\label{fig:3}
\end{figure}

We further measured the angular dependences of interfacial Rashba MR for the three rotations at various back gate voltages [Fig. \ref{fig:4}(a)].
In the $\alpha$ ($\beta$)-rotation, the interfacial Rashba MR in general follows $\cos^2 \alpha$ ($-\cos^2 \beta$), consistent with Eq.~\eqref{eq1}.
An exception appears for $V_g$ = 20 V. At this gate voltage, $R_{xx}$ in the $\alpha$-rotation is asymmetric between 90 and 270 degrees.
This asymmetry is attributed to a nonnegligible contribution from current-induced spin polarization as reported previously~\cite{Narayanapillai}.
For detailed description of the current-induced spin polarization to the magneto-resistive effect, see Ref.~[\onlinecite{Ganichev}].
In the $\beta$-rotation, on the other hand, the sign of interfacial Rashba MR changes at $V_g$ = 20 V. This sign change is attributed to the WL correction as the $\beta$-rotation involves the contribution of out-of-plane component of the magnetic field $B_{\rm ext}$ and the WL correction to $R_{xx}$ becomes significant at low gate voltages (see Fig. \ref{fig:2}).


\section{Theoretical analysis}

In order to understand the interfacial Rashba MR, we compute charge transport in the 2DEG at the LAO/STO interface.
As an effective model which captures the qualitative features of the experimental results, we choose a simple model Hamiltonian $\cal H$ as
\begin{align}
{\cal H}(\v k) = \frac{\hbar^2 {\v k}^2}{2 m} + \alpha_R \boldsymbol{\sigma}\cdot (\hat {\v z} \times \v k) + J \boldsymbol{\sigma}\cdot {\hat {\bf m}},\label{Hamil}
\end{align}
where $\v k$ [= ($k_x$, $k_y$)] is a two-dimensional wave vector, $m$ is the effective mass of the $d_{xy}$ band, $\hbar$ is the reduced Planck constant, $\alpha_R$ is the Rashba constant, $J$ is the exchange coupling between the conduction electron spin $\boldsymbol{\sigma}$ (Pauli matrices) and unit localized magnetic moment $\hat {\bf m}$.
Here we assume that the magnetization orientation ($\hat {\bf m}$) is aligned along the external magnetic field direction.
In general, the Rashba constant depends on the magnetization orientation~\cite{Gmitra, Chen2016}. In our model calculation, we ignore this angular dependence of the Rashba constant and show that qualitative features of the angle-dependent MR observed in experiment are described by a simple free electron model without considering additional angular dependence of physical properties.
The choice of this Hamitonian demands an explanation. Since the STO is a cubic perovskite, three $t_{2g}$ orbitals are degenerate at the $\Gamma$ point in the bulk STO. For STO-based interface, the $d_{xy}$ band lies lower than the degenerate $d_{yz}$ and $d_{xz}$ bands due to the interface confinement~\cite{Kim2013,Zhong,Khalsa}. The 2DEG geometry leads to a circular Fermi surface for $d_{xy}$ band and degenerate elliptical Fermi surfaces for $d_{yz}$ and $d_{xz}$ bands. In our experiment, the sample is conducting at high gate voltages and becomes insulating as the gate voltage decreases.
Moreover, the $\alpha$-rotation results of $R_{xx}$
[Fig. \ref{fig:4}(a)] show $\cos^2\alpha$-like dependence without the Lifshitz transition which was observed in Ref~[\onlinecite{Joshua}]. These results indicate that the Fermi level lies in the the lowest $d_{xy}$ band and lowers as the gate voltage decreases. Therefore, in our theoretical analysis, we use a free-electron Hamiltonain with Rashba SOC and $s$-$d$ exchange interaction by focusing on the single $d_{xy}$ band. For simplicity, we treat the vector potential contribution to MR separately as the WL correction.

In our model calculation, we focus on the qualitative description of the longitudinal conductivity.
In the high gate voltage regime where the WL correction is subdominant, the qualitative feature of the longitudinal conductivity
is captured by the Kubo formula with the relaxation time approximation, given as
\begin{align}
&\sigma_{xx}^0 = 2 e^2 \tau \sum_{n=\pm} \int \frac{d^2 k}{(2\pi)^2} \left(v^n_x\right)^2 \delta\left(E_n -E_F\right),\\
&E_\pm=\frac{\hbar^2 {\v k}^2}{2m}\pm \left|\alpha_R (\v k \times \hat {\v z}) + J {\hat {\bf m}} \right|,\\
&v_x^\pm = \frac{\hbar k_x}{m}\pm \frac{\alpha_R}{\hbar}\left(\frac{J m_y + \alpha_R k_x}{\left|\alpha_R(\v k \times \hat {\v z})+J {\hat {\bf m}}\right| }\right),\label{vel}
\end{align}
where $+/-$  represents spin up/down band, $e$  is the electron charge, $\tau$ is the relaxation time, and $E_F$ is the Fermi energy.
In the experimental result depicted in Fig. 4(a), there are sign changes between low gate and high gate voltage regimes.
In order to describe the electron transport in the low conductance (i.e., low gate) regime,
we adopt the theoretical results of Ref.~[\onlinecite{Iordanskii,Knap}], which compute the WL correction as
\begin{widetext}
\begin{align}\label{eqWL}
\Delta \sigma_{xx}(B) = -\frac{e^2}{4\pi^2 \hbar} \Bigg\{&\frac{1}{a_0}+\frac{2 a_0 +1 + H_{SO}/B}{a_1\left[a_0 + H_{SO}/B\right]-2H_{SO}/B}+2\,{\rm ln}\frac{H_{tr}}{B}+\psi\left(\frac12 + \frac{H_\phi}{B}\right)+3C\nn
&-\sum_{n=1}^{\infty}\left[\frac{3}{n}-\frac{3a_n^2 + 2a_n H_{SO}/B - 1 - 2(2n+1)H_{SO}/B}{\left[a_n + H_{SO}/B\right]a_{n-1}a_{n+1}-2(H_{SO}/B)\left[(2n+1)a_n - 1\right]}\right]\Bigg\},
\end{align}
\begin{figure}[ttbp]
\includegraphics[width=170mm]{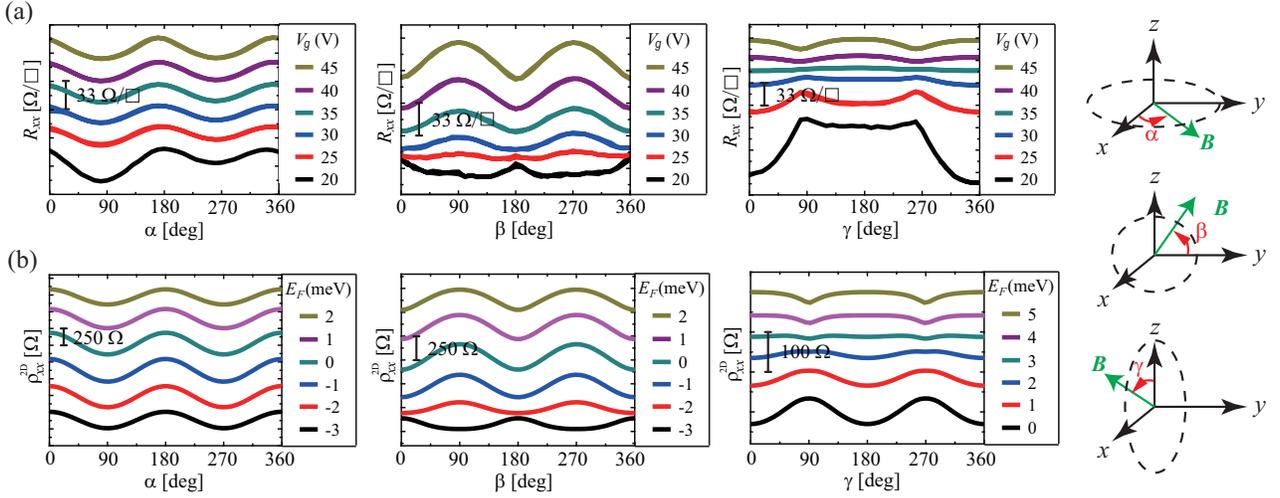}
\caption{(color online) (a) Experimental results of angular dependence of $R_{xx}$ as a function of $\alpha$-, $\beta$-, and $\gamma$-rotations with various $V_g$, and applied field $B_{\rm ext}=9$ T.
(b) Theoretical calculations for angular dependence of the resistivity, $\rho^{2D}_{xx}$, as a function of $\alpha$-, $\beta$-, and $\gamma$-rotations with various Fermi energies. 
The curves are offset along the $y$-axis for clarity.
}\label{fig:4}
\end{figure}
\end{widetext}
where $a_n = n+\frac{1}{2} + \frac{H_\phi}{B}+ \frac{H_{SO}}{B}$, $\psi(1+z) = -C + \sum_{n=1}^{\infty}\frac{z}{n(n+z)}$ and $C$ is the Euler constant.
We note that Eq. \eqref{eqWL} is a quantum correction to the conductivity due to a perpendicular magnetic field~\cite{Bin}. We use the following parameters for the model calculations: $m=0.7 m_e$ ($m_e$ is the free electron mass)~\cite{Hurand},  $J=2.5$ meV, $\tau=0.33$ ps, $H_\phi=3.0\,$T, and $H_{SO}=1.2\,$T. For gate voltage effect on the Rashba coefficient, we assume $\alpha_R =(60 + \lambda\, E_F)\,${\rm meV}$\rm {\AA}$, where $\lambda = 4/{\rm meV}$.

In Fig. \ref{fig:4}(b), we show theoretical results for the angular dependence of the resistivity as functions of the rotating angles in $\alpha$, $\beta$, and $\gamma$. The change in the gate voltage is considered as the change in the Fermi energy. The Fermi-energy shift for the $\gamma$-rotation is taken into account in order to reflect the resistance hysteresis from the voltage cycle~\cite{note1}. In all three rotations, the theoretical calculations of MR qualitatively matches well with the experimental results [Fig. \ref{fig:4}(a)].
Even though our model Hamiltonian and calculation scheme is simplified,
this good agreement shows that the simple Hamiltonian [Eq.~\eqref{Hamil}] considering the coexistence of Rashba SOC and exchange coupling describes the experimental result reasonably well and the interfacial Rashba SOC is key to the interfacial Rashba MR of which angular dependence is similar to that of the spin Hall MR.


In order to get an insight into the similarity between the interfacial Rashba MR and the spin Hall MR, we focus on the second term in Eq.~\eqref{vel}, which is an additional velocity originating from the Rashba interaction. This additional velocity includes the magnetization orientation that gives the angular dependence of the longitudinal conductivity. In other words, a strong $m_y$ dependence of $\Delta R_{xx}$ is natural because of the symmetry of Rashba SOC. For instance, one obtains $v_x^\pm \approx \hbar k_x/m \pm J m_y \alpha_R/\hbar$ for $J\gg\alpha k_F$.
There is also an additional source of the angular dependence of the interfacial Rashba MR. When the Rashba SOC and exchange coupling coexist, the Fermi surface distorts depending on the magnetization direction~\cite{Lee2013}. The Fermi surface distortion is maximized for an in-plane magnetization whereas it is absent for an out-of-plane magnetization. Because of these two contributions, additional velocity and Fermi surface distortion, the interface subject to both Rashba and exchange interactions exhibits the interfacial Rashba MR similar to the spin Hall MR.

\section{Summary and Discussion}

In summary, we report the interfacial Rashba MR for LAO/STO 2DEG and find that its angular dependence is similar to that of the spin Hall MR observed in FM/HM structures. Our model calculations describe the experimental results reasonably and show that the spin-Hall-MR-like behavior originates from the combined effects of the interfacial Rashba SOC and exchange coupling. As the bulk spin Hall effect is absent in the LAO/STO 2DEG system, our finding evidences that the interfacial Rashba SOC gives rise to the spin-Hall-like MR. Therefore, our result suggests that the interfacial SOC effect is key to understanding of various SOC-related phenomena in magnetic/non-magnetic bilayers.

In the theory part of this work, we note that there are potential complexities left aside.
In this paper, we use the free electron model of single orbital ($d_{xy}$) with the linear Rashba interaction.
However, the LAO/STO 2DEG is composed of $t_{2g}$ orbitals with nonquadratic energy dispersion and there is a report suggesting the existence of the cubic Rashba interaction~\cite{Nakamura}.
Moreover, in our model, many parameters are difficult to be determined in experiment and a quantitative description including the quantum correction is not reliable.
Thus, we restrict the purpose of model calculation to a qualitative description of the experimental observation.

We end this paper by noting that a recent experiment for a Bi/Ag/CoFeB metallic trilayer found the Rashba-Edelstein MR~\cite{Nakayama2016}, of which angular depedence is similar to that of the spin Hall MR. This Rashba-Edelstein MR originates from the combined action of two separate interfaces, the Bi/Ag interface with Rashba SOC~\cite{Ast} and the Ag/CoFeB interface with the exchange splitting, through a spin diffusion process. In this trilayer including a conducting bulk, therefore, the charge-to-spin and spin-to-charge conversions at the Bi/Ag interface and the spin-dependent reflection at the Ag/CoFeB interface are separated. Because of this separation in the trilayer, it is not straightforward to get an insight into the bulk versus interface contributions to the spin Hall MR of the FM/HM bilayers. In contrast, our result gives a clear indication about the pure interface contribution to the spin Hall MR as the LAO/STO 2DEG, just like the FM/HM interface, is a single interface subject to both Rashba SOC and exchange coupling with no conducting bulk.

\begin{acknowledgments}
This work was supported by the National Research Foundation (NRF), Prime Ministers Office, Singapore, under its Competitive Research Programme (NRF CRP12-2013-01), the National Research Foundation of Korea (NRF-2015M3D1A1070465, NRF-2016R1A6A3A11935881, NRF-2017R1A2B2006119), and by the DGIST R\&D Program of the Ministry of Science, ICT and Future Planning (17-BT-02).
\end{acknowledgments}


\begin{thebibliography}{99}
\bibitem{note1} We note that for the MR measurement, we sweep the field and gate voltage for a given angle-rotation.
Because of the resistance hysteresis originating from the voltage cycle, the MR of an angle-rotation is not the same as the sum of MRs of other two angle-rotations.
\bibitem{Bin} For simplicity, we neglect the quantum correction from an in-plane magnetic field
which is small compared to that from a perpendicular magnetic field at high field regimes (see Ref.~[\onlinecite{Maekawa,Komori}]).
\end{thebibliography}
\end{document}